\pdfoutput=1
\documentclass[onecolumn,preprintnumbers,amsmath,amssymb]{revtex4}

\usepackage{graphicx}
\usepackage{dcolumn}
\usepackage{bm}

\begin{document}

\preprint{APS/123-QED}

\title{RISK OF POPULATION EXTINCTION FROM PERIODIC AND ABRUPT CHANGES OF ENVIRONMENT}


\author{Andrzej Pekalski}
\email{apekal@ift.uni.wroc.pl}
\affiliation{Institute of Theoretical Physics, University of Wroclaw,\\
 pl. M. Borna 9, 50-203 Wroclaw, Poland}
\author{Marcel Ausloos}
\email{Marcel.Ausloos@ulg.ac.be}
\affiliation{GRAPES@SUPRATECS B5a Sart Tilman, B-4000, LIEGE, Euroland}


\begin{abstract}
A simulation model of a population having internal (genetic)
structure is presented. The population is subject to selection
pressure coming from the environment which is the same in the whole
system but changes in time. Reproduction has a sexual character with
recombination and mutation. Two cases are considered - oscillatory
changes of the environment and abrupt ones (catastrophes). We show
how the survival chance of a population depends on maximum allowed
size of the population, the length of the genotypes characterising
individuals, selection pressure and the characteristics of the
''climate`` changes, either their period of oscillations or the
scale of the abrupt shift.
\end{abstract}


\maketitle

\section{Introduction}
 Estimation of the extinction risk for a population is, obviously, an important issue. It has been addressed
in many papers, either by biologists (see e.g
\cite{lande}-\cite{shannon}) or physicists (\cite{higgs}
-\cite{gang4}). Several aspects have been considered -- most often
it was the problem of changing environment, like advancing ice-age
\cite{pease,mdap} and the question was -- will the population adapt,
or migrate? The effect of stochastic changes and random catastrophes
on the population's fate has been studied, via mean-field type
analysis, by Lande \cite{lande}. Roberts and Newman \cite{roberts}
studied an extension of the Bak and Sneppen model \cite{bs}, taking
into account both bad genes and bad luck, represented by a
catastrophe. In most of the papers describing population dynamics
the genetic structure of the population has not been considered.
Individuals were
characterized by their continuous trait, represented by a real number
 $z_i \in$ [0,1], see e.g. \cite{burger,gang4}. Although in many cases
 such simplified approach is quite satisfactory, it cannot describe,
 for example, the influence of the genetic structure on the survival probability.
  In more refined models, individuals are characterized only be their genotypes
   \cite{higgs,hall} which are subject to random mutations. Individual-based
   model of evolution has been recently proposed by Rikvold and Zia \cite{rikvold}.
    They used a fixed in time and random interaction matrix characterizing species and their phenotypes. The latter could be changed by random mutations. The model exhibits punctuated equilibrium -- short periods with many changes in the genome space, separated by long periods of stassis.\\
In this paper we study how the survival chance of a population depends on such factors as type of the
environmental changes (oscillatory or abrupt), length of the genotype characterising individuals belonging to
the population and selection pressure. We shall use Monte Carlo (MC) simulations of a discrete time model.

\section{Model}

In our model a population is, at time $t$, composed of $N(t)$ individuals, which have no spatial location and
are described by their age, which is increasing after each time step (see below) and their genotypes. A
genotype consists of a double string (the organisms are diploidal) of  $L$ sites (loci) equal either zero or
one.
From a genotype a phenotype (single string) is constructed by taking at each site the product of the two
values on both strings of the genotype. Hence the phenotype is also composed of a zeros and ones
\cite{fraser}. For $L$ = 5 the process could be illustrated as follows\\[3ex]
\hspace*{6mm} Genotype \hspace*{55mm} Phenotype\\[2ex]
\hspace*{3mm} 0 \quad 1 \quad 1 \quad 0 \quad 1 \\[-1ex]
\hspace*{55mm} $\Rightarrow$ \hspace*{10mm} 0 \quad 0 \quad 1 \quad 0 \quad 0\\[-1ex]
\hspace*{3mm} 0 \quad 0 \quad 1 \quad 1 \quad 0\\[2ex]
As can be seen, $0$ is the dominant and $1$ is the recessive allele. The population lives in a habitat which
is characterized by an optimum \cite{burger}, $\Theta$(t), which is, like the phenotype, a single string of
$0$'s and $1$'s of length $L$. The agreement between the optimum and  an individual $i$ phenotype $f_i$
determines its fitness $\varphi_i$
\begin{equation}
\label{fitn}
\varphi_i \,=\, \frac{1}{L} \sum_{i=1}^L \left[1 - XOR(f_i,\Theta(t))\right].
\end{equation}
where $XOR$ is the exclusive $OR$, equal 1 if either $f_i$ or $\Theta$  is equal 1, but not both.
Therefore an individual having a phenotype equal to the optimum has the maximum probability of survival.
This probability, $p_i$, is calculated from \cite{burger,mdap}
\begin{equation}
\label{prob}
p_i\,=\, \exp\left( - \frac{s \cdot w_i}{\varphi_i}\right),
\end{equation}
where $s$ is the selection pressure which may describe how demanding is the environment. The larger is $s$,
the more demanding is the habitat.  An individual with a given fitness is less likely to survive when the selection pressure is high, since its survival probability is smaller, than when the selection pressure $s$ is small.  $w_i$ is the age of the individual $i$. Initial values of the age and genotypes are random.

Our MC simulations follow the steps given below.
\begin{enumerate}
\item Pick the first available individual from a list,
\item Its fitness is calculated from eq.(\ref{fitn}), then probability of survival, $p_i$ from eq.(\ref{prob}),
\item A random number, $r_i \in$ [0,1] is taken from a uniform distribution. If $r_i > p_i$ then the
individual is removed from the system and the program goes back to 1.
\item If the individual survived, the next one is taken from the list as a partner for reproduction, and its
survival probability is checked, like for the individual $i$. If it did not survived, again the program
returns to 1,
\item The pair gives birth to up to 4 offspring. That value has been chosen since for a smaller one, say 2,
the populations will soon die out, and larger values, like 6, will not change the results in any significant way.
Each of the offspring is born if a random number $r \in$ [0,1] is smaller
than the Verhulst factor \cite{verhulst}
$$
\pi = 1 \, - \frac{N(t)}{K},
$$
where $K$ is the maximum number of individuals the habitat could
support (carrying capacity).  Therefore the Verhulst factor could be
regarded as an yet another factor, apart from the selection
pressure, limiting the growth of a population. The difference
between the two is that the Verhulst factor acts only on offspring,
which are either born or not. Because of the Verhulst factor the
number of litter at a given birth could be any integer number
between 0 and 4,
\item Each progeny receives its genotype via recombination and mutation \cite{fraser,mdap}. The two strings
of the first parent's genotype are cut at a random position and then
glued across. From the two one string (a gamete) is chosen randomly
and in one position the allele is changed (mutated) to the opposite.
The chosen gamete will be one of the two chromosomes of the genotype
of the offspring. The second chromosome is obtained from the second
parent, following the same steps.  From the genotype the phenotype
is constructed in the way described above. Since the place of
cutting the strings,  mutated locus and the choice of the gametes,
are random, each offspring coming from the same parents may have a
different genotype,
\item After coming to the end of the list of individuals, the list is updated and shuffled. The time step as
well as the age of the individuals is increased by one. The reason to include age is to get rid of perfectly
fit individuals who otherwise would live forever in a constant environment.
 \end{enumerate}
A population in which a partner is chosen freely from all members of the population is called {\it panmictic}.

We shall consider below two cases. In each of them initially the optimum will be a string of zeros. Since
zero is the dominant allele, this corresponds to a ''friendly`` climate (three combinations of alleles in a
genotype yield a zero in the phenotype, while only one combination gives $1$). All $1$'s in
the optimum mark the most ''harsh`` climate. In the first of the cases the optimum will change periodically,
with a period denoted by $t_{ch}$, while in the second case it will change just once, after the system
reached a stationary state. The degree of changes in this case will be measured by the number $b$ of zeros in
the optimum switched from 0 to 1's.

Our model has the following control parameters: maximum size of the system $K$, length of the
genotypes $L$, selection pressure $s$ and either period of changes $t_{ch}$ or the number of bits $b$ changed
in the optimum.

Typically we have run the simulations till 10 kMCS and averaged over 50 independent runs for larger systems
and 500 for smaller ones. Time of extinction was determined as that moment when there was just one individual
left in the population. Survival chance for a population was determined as the ratio of the number of runs in which a population survived to the end of simulations to the total number of runs.

\section{ Results}
\subsection{Oscillations of the optimum}
The behavior of populations in an oscillating environment has been
recently  studied by mean-field analysis and simulations in
\cite{gang4}, where however no genetic structure has been considered
and the populations were living on a lattice.

\begin{figure}
\begin{center}
\includegraphics[scale=0.8]{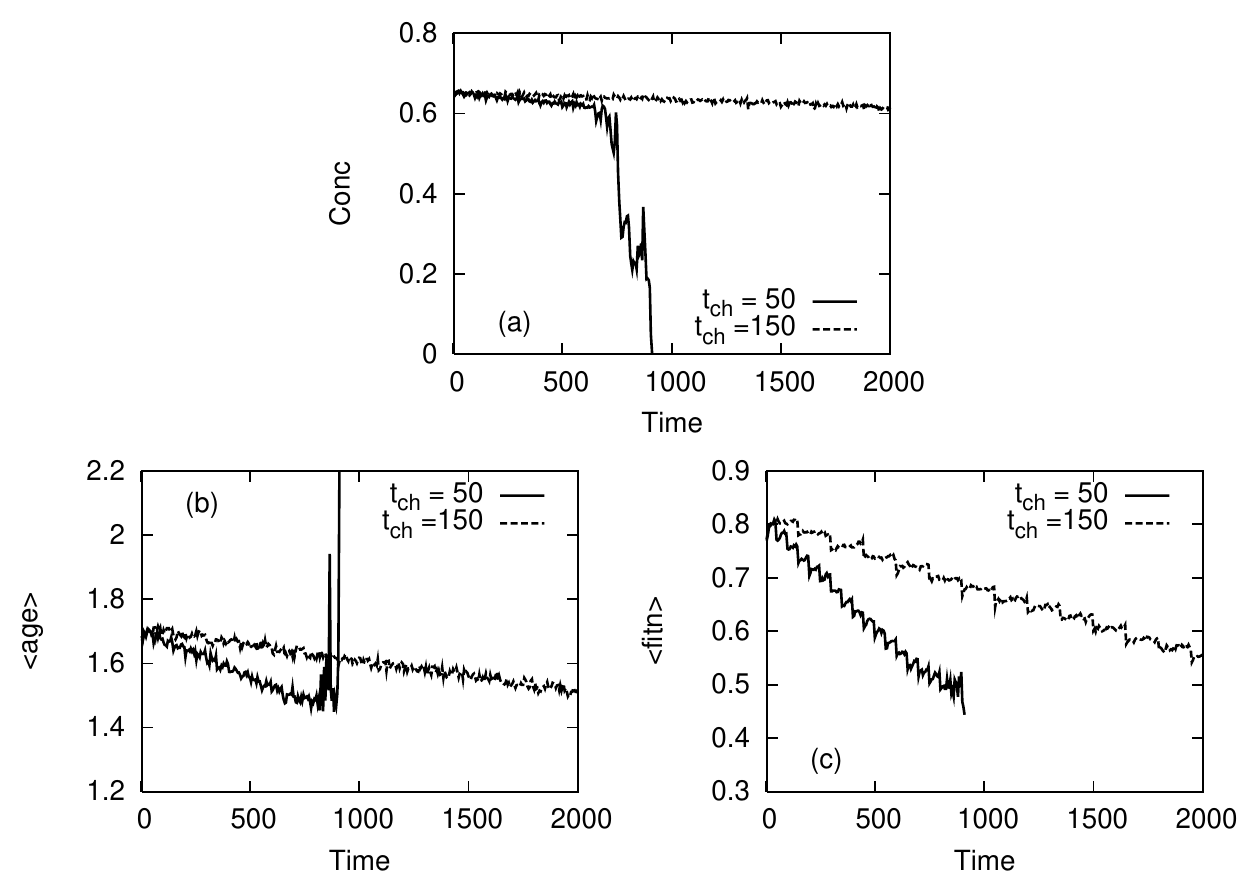}
\caption{Time dependence of (a) concentration, (b) average age and
(c)  average fitness, when the optimum changes with periods $t_{ch}$
= 50 MCS and $t_{ch}$ = 150 MCS. Carrying capacity $K$ = 2500,
genotype length $L$ = 20, selection pressure $s$ = 0.17, average
over 50 runs. }\label{fig1}
\end{center}

\end{figure}

\begin{figure}
\begin{center}
\includegraphics[scale=0.5]{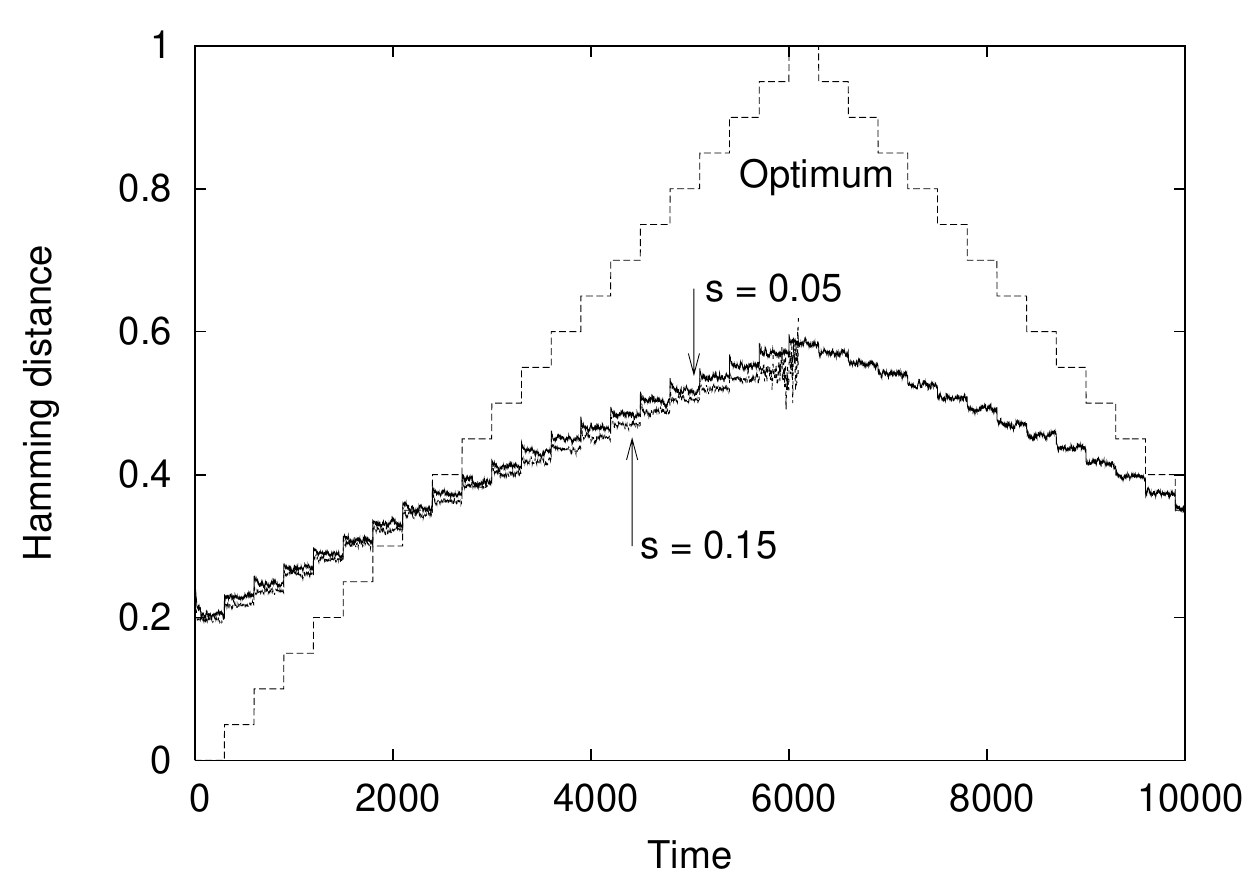}
\caption{Average Hamming distance between optimum and a phenotype in
the case of population surviving and vanishing. $t_{ch}$ = 300 MCS
}\label{fig2}
\end{center}

\end{figure}

In our model the optimum was changed with periods of $t_{ch}$ = 20,
30, 50, 70, 100 and 150 time units (MCS). In Figure \ref{fig1} we
show the time evolution of the concentration, average age and
average fitness for fast ($t_{ch}$ = 50) and slow ($t_{ch}$ = 150
MCS) changes. As can be seen, the populations go extinct much sooner
for faster changes of the optimum, what have been also found in
\cite{gang4,burger}. Average fitness oscillates following the
optimum but diminishes rather fast, indicating that populations
could not adapt to the changing conditions. Relatively stable
concentration is maintained due to a large number of offspring,
which shows up in decreasing average age. Population enters into a
critical region when the number of individuals is so low that a
chance to meet a partner and to breed is smaller than the average
survival probability. Progeny is not born, the average age jumps up
and the fitness continues to drop. Since the average age is about
1.7, even a fast change of the optimum, like $t_{ch}$ = 20,
corresponds to about 15 generations, while
$t_{ch}$ = 150 is about 100 generations. In our model the generations
 are overlapping, meaning that parents do not die after giving birth to
 offspring. We have found out that a population could either adapt
 to the changing conditions, or go extinct. Since however the optimum
 does not change in space, we cannot have islands serving as a refuge
 for otherwise declining population, as has been found out in \cite{shnerb}.\\
Simple following of the optimum by the average phenotype may not be
however a guarantee of survival for a population, as seen from
Figure \ref{fig2}, where average Hamming distance \cite{drossel}
between the optimum and the phenotypes is shown for  slow changes of
the optimum ($t_{ch}$ = 300) and two values of the selection -- low
($s$ = 0.05) and high ($s$ = 0.15). Although the Hamming distance is
smaller for high selection, i.e. the phenotypes follow more closely
the optimum, populations could not sustain the high killing rate,
eliminating a wider range of ill-fitted individuals, and they are
wiped out. Populations can not  survive very difficult conditions,
represented by optimum close to $1$. Let us remind here that a $1$
in a phenotype comes only from one pair (1,1) of alleles on the two
chromosomes, while the remaining three combinations (0,0), (0,1) and
(1,0) produce a $0$ in the phenotype. This probabilistic elimination
of individuals with low fitness agrees  with the Darwinian {\it
survival of the fittest } and the mechanisms described by Roberts
and Newman \cite{roberts}, who considered also the effect of ''bad
luck''.

\begin{figure}
\begin{center}
\includegraphics[scale=0.5]{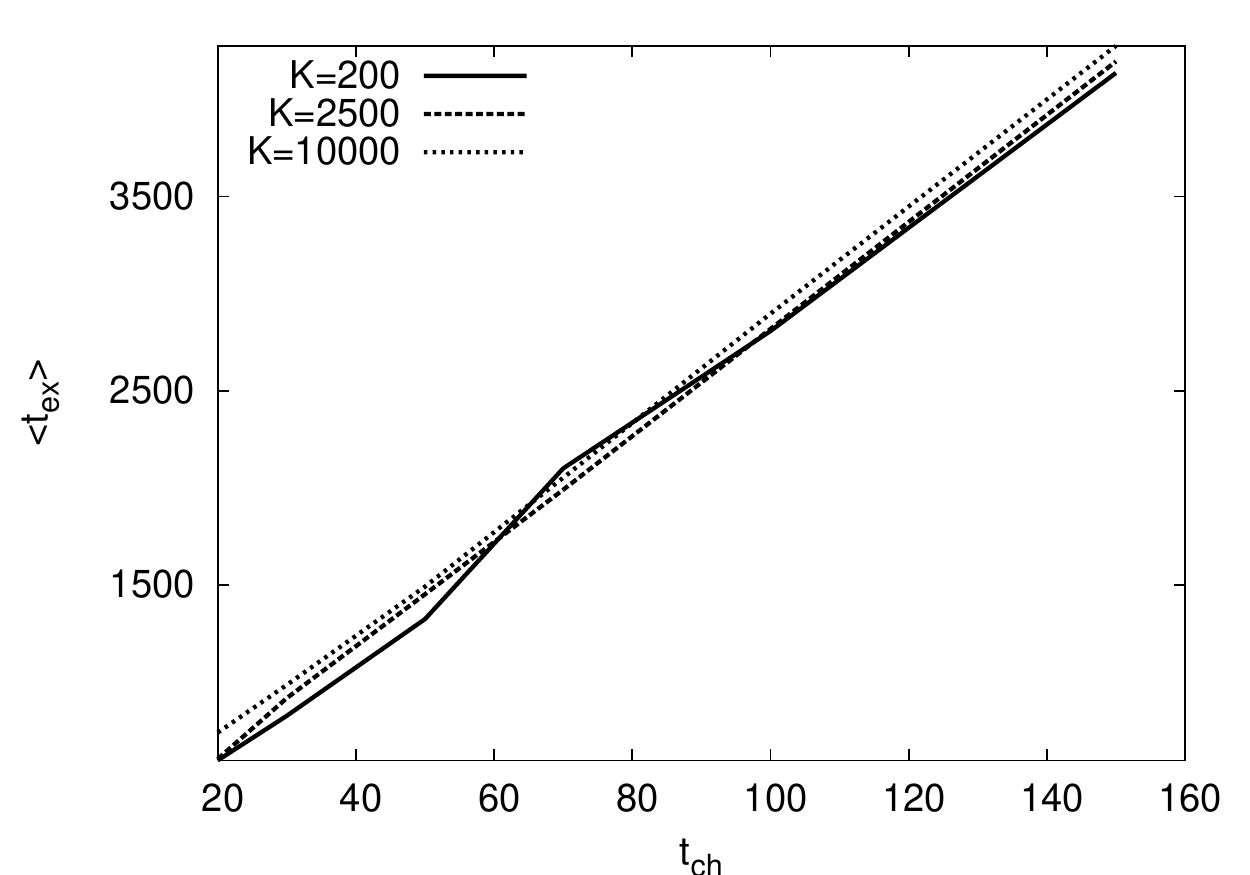}
\caption{Average time to extinction $<t_{ex}>$ versus periods
$t_{ch}$   of optimum oscillations for small ($K$ = 200), medium
($K$=2500) and large ($K$ = 10000) carrying capacities. $L$ = 30,
$s$ = 0.16, except for $K$= 200, where $s$ = 0.15 } \label{fig3}
\end{center}

\end{figure}

The dependence of the average time of extinction, $<t_{ex}>$ on the
period of changes, $t_{ch}$, is shown in Figure \ref{fig3}, which
also demonstrates the dependence of $<t_{ex}>$ on the size of the
populations. As could be expected, there exists a minimum value of
the selection for which all populations died out. For smaller
selections some populations would survive. This threshold value of
the selection will be henceforth denoted by $s_c$. The results shown
below are  for the threshold values equal
 $ s_c$ = 0.16  for $K$ = 2500 and  $K$ = 10000 and for $s_c$ = 0.15 for $K$ = 200.
 We have observed here a well known fact \cite{shaffer,shnerb2} that small populations are more
vulnerable and a weaker selection pressure drives them to extinction.\\
As seen, average extinction time increases linearly with $t_{ch}$ and the slope is practically independent of
the maximum size of the population.

\begin{figure}
\begin{center}
\includegraphics[scale=0.5]{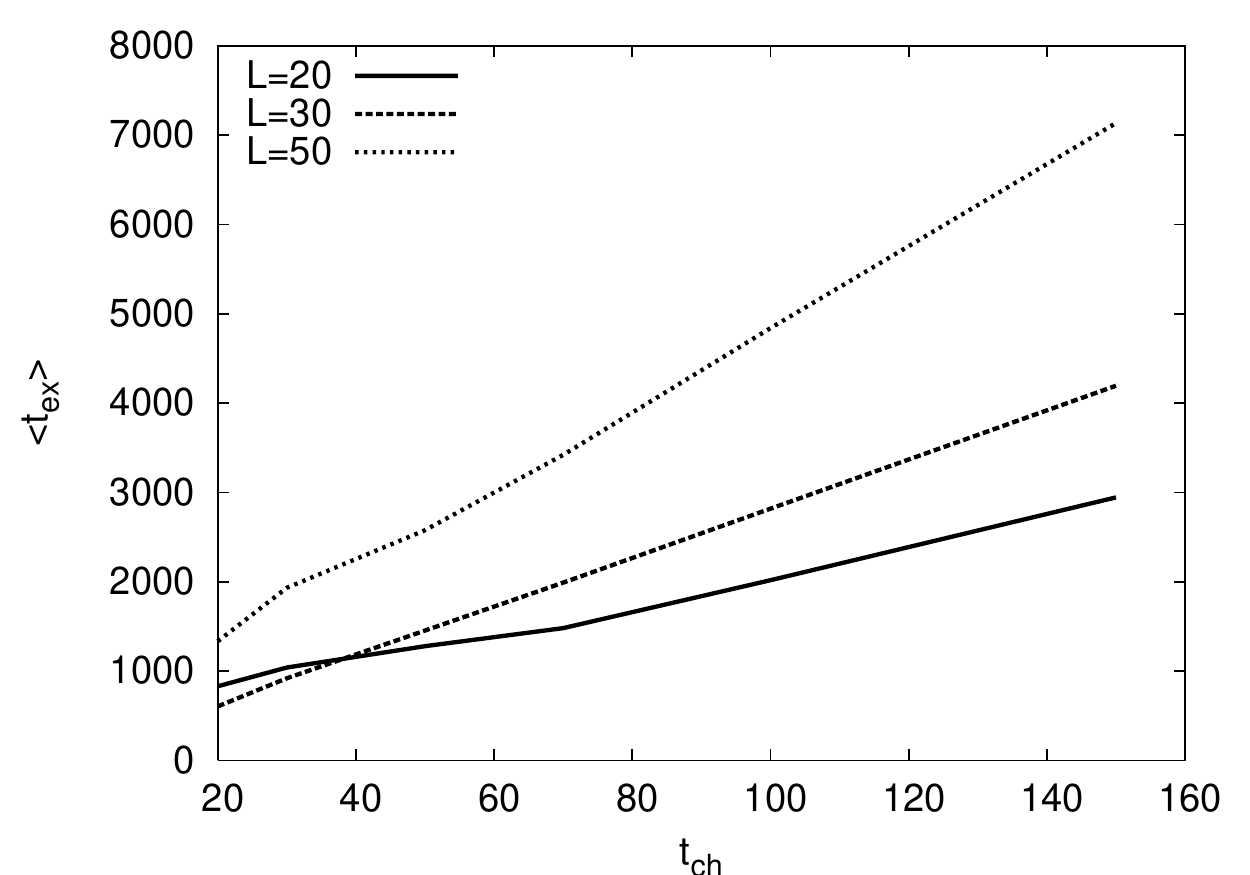}
\caption{Average time to extinction $<t_{ex}>$ versus periods
$t_{ch}$  of optimum oscillations for three values of the genotype
length -- $L$ = 20, 30 , 50. Carrying capacity $K$ = 2500, $s$ =
0.16 } \label{fig4}
\end{center}
\end{figure}

For selections stronger than $s_c$ we observe also linear dependence
of $<t_{ex}>$ on $t_{ch}$, with the same slope, but lying lower than
for $s_c$. Figure \ref{fig4} shows how $<t_{ex}>$ depends on
$t_{ch}$ when the length of the genotype changes. The maximum size
of the system was $K$ = 2500. Clearly, individuals with longer
genotypes (more complex) are better off, live longer, than the ones
with shorter genotypes. In each case we have observed a linear
dependence of $<t_{ex}>$ on $t_{ch}$.

\begin{figure}
\centering
\includegraphics[scale=0.5]{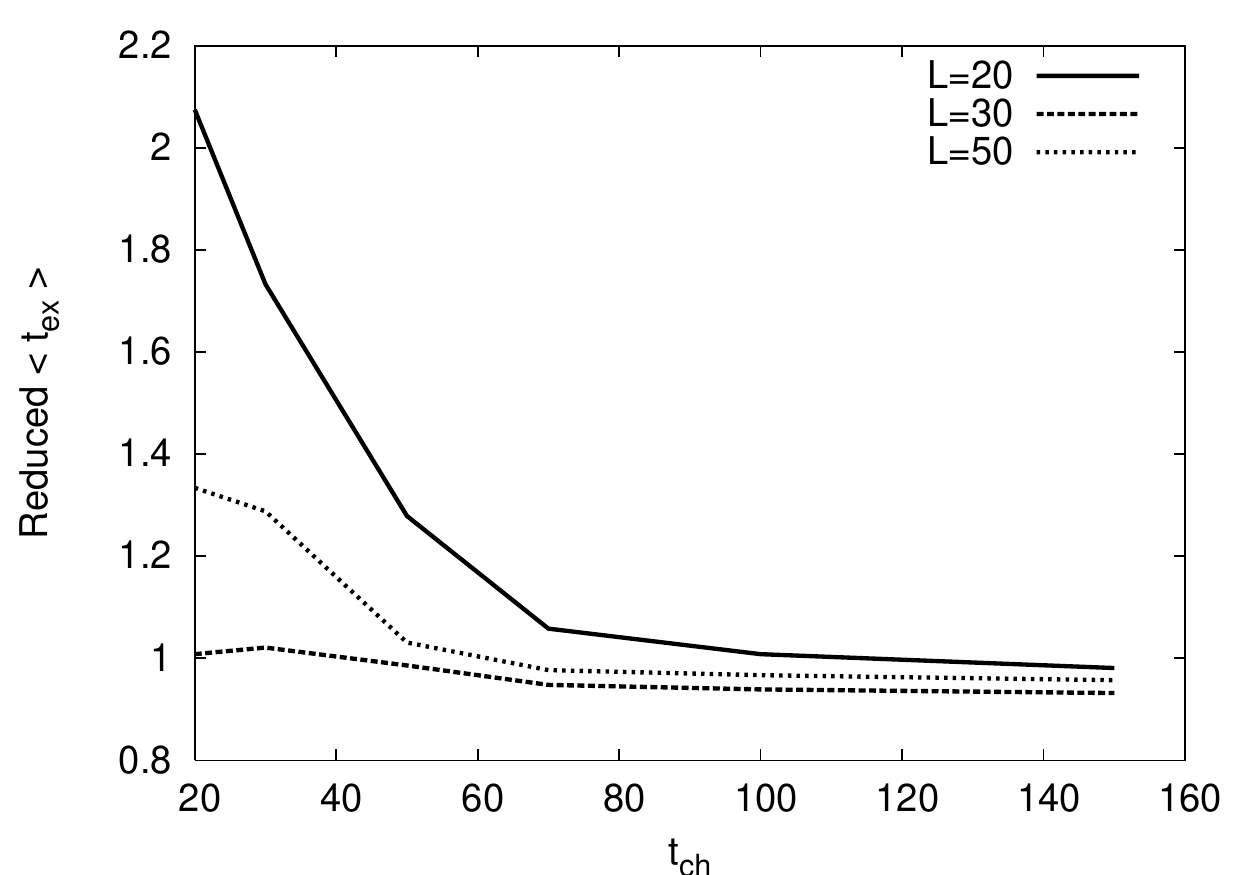}
\caption{Reduced average time to extinction (see text) versus
$t_{ch}$. Parameters' values are the same as in Figure \ref{fig4}. }
\label{fig5}
\end{figure}

Average extinction time reduced by the product of the period of the
changes and the length of the genotype is shown in Figure
\ref{fig5}. While for fast changes we observe differences among
various cases, for long-period oscillations the reduced $<t_{ex}>$
stabilizes at about 1.

\begin{figure}
\centering
\includegraphics[scale=0.5]{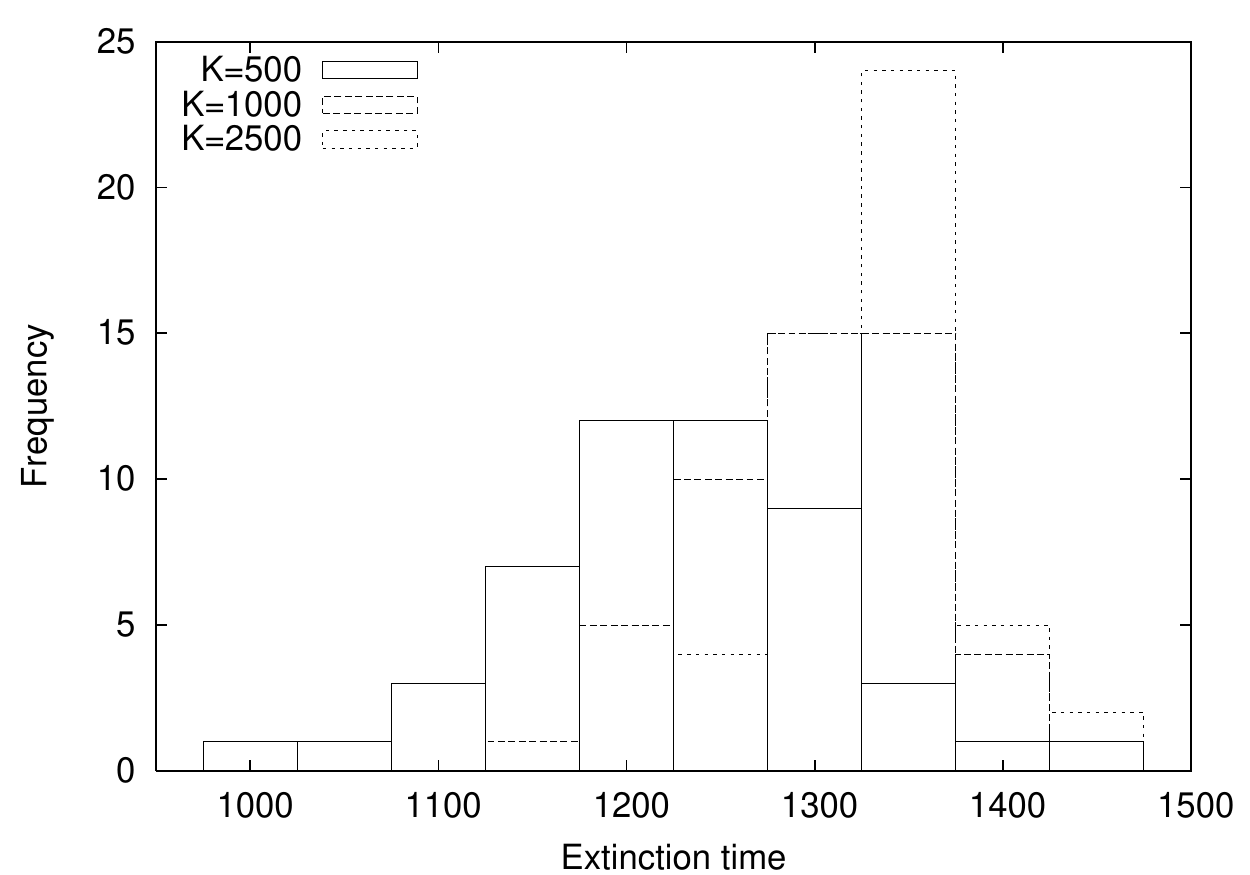}
\caption{Distribution of the extinction times at 50 independent runs
for different carrying capacity values $K$. $L$ = 30, $s$ = 0.16,
$t_{ch}$ = 50.} \label{fig6}
\end{figure}

To ensure that the most often used averaging over just 50
independent runs yields good statistics, we present in Figure
\ref{fig6} the values of $<t_{ex}>$  obtained in 50 runs for $L$ =
30, $t_{ch}$ = 50, $s$ = 0.16 and several values of $K$. Apart from
very small populations ($K$ = 200), all other systems show rather
small scatter. Therefore in the following for $K$ = 200 we took
averages over 500 runs.

It should be noticed that the survival chance of a population
depends very strongly on the selection pressure. There is a range of
the selection pressure values within which some populations may die,
while some may stay alive. Outside that range either all populations
die or all stay alive. For example for $K$ = 2500, $L$ = 50,
$t_{ch}$ = 50 at $s$ = 0.14 all populations survive, meaning that
they were able to adapt (continuously) to the changing habitat. At
$s$ = 0.15 only 15 \% survive, and at $s$ = 0.16 all die. Survival
chance as a function of the selection pressure has a nearly
step-like character. The threshold values of the selection, $s_c$,
are equal 0.16 for $L$ = 50 and $L$ = 30, irrespective of the rate
of changes $t_{ch}$ and $s_c$ = 0.15 for $L$ = 20. This means that
populations of individuals with shorter genotypes are more
vulnerable than those with longer ones. In general, the system
always tries to follow the optimum. If the selection is too strong,
then the distance between the average phenotype and the optimum is
small, but many individuals are killed  and the killing rate may be
too high for the population to survive. If the selection is weaker,
the distance is larger, but less individuals are killed and the
population survives.
\subsection{Abrupt changes of the optimum}
\begin{figure}
\centering
\includegraphics[scale=0.8]{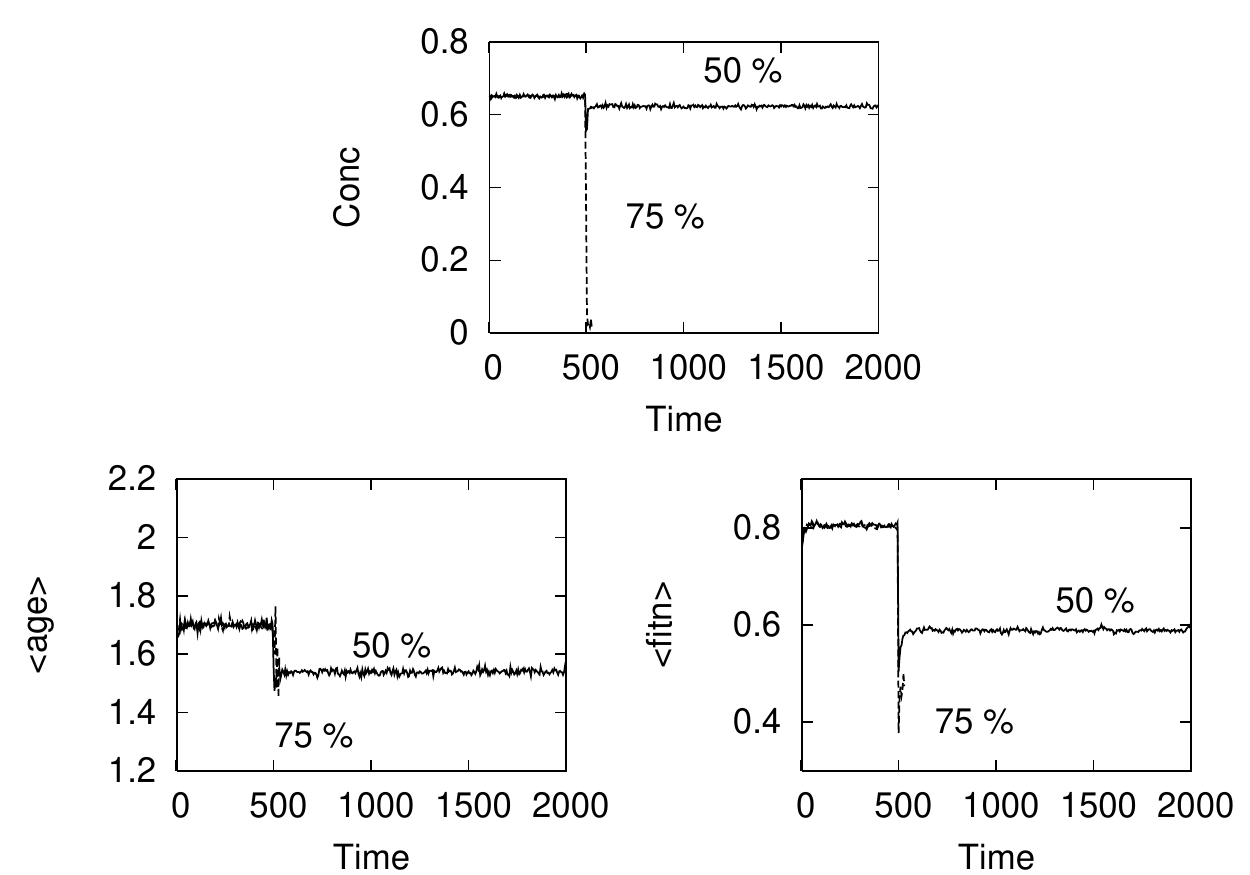}
\caption{Time dependence of (a) concentration, (b) average age and
(c) average fitness when the environment changed after 500 MCS. Two
types of changes -- $b$ = 50 \%  and  $b$ = 75 \%  (loci in the
optimum changed). $K$ = 2500, $L$ = 20, $s$ = 0.15} \label{fig7}
\end{figure}

Let us present the time dependence of the
concentration, average fitness and average age (Figure \ref{fig7}).

The system shown is a medium size population ($K$ = 2500) with a
short genotype ($L$ = 20). We let it evolve in a constant
environment until 500 MCS when the population reached a stationary
state, and then changed either half of the zeros in the optimum to
ones, or 75 percent of zeros to ones. Afterwards the optimum
remained constant, but with the new values. If a population survived
the shock of the change, it will continue to exist, although with
lower average fitness, and lower average age.

\begin{figure}
\centering
\includegraphics[scale=0.8]{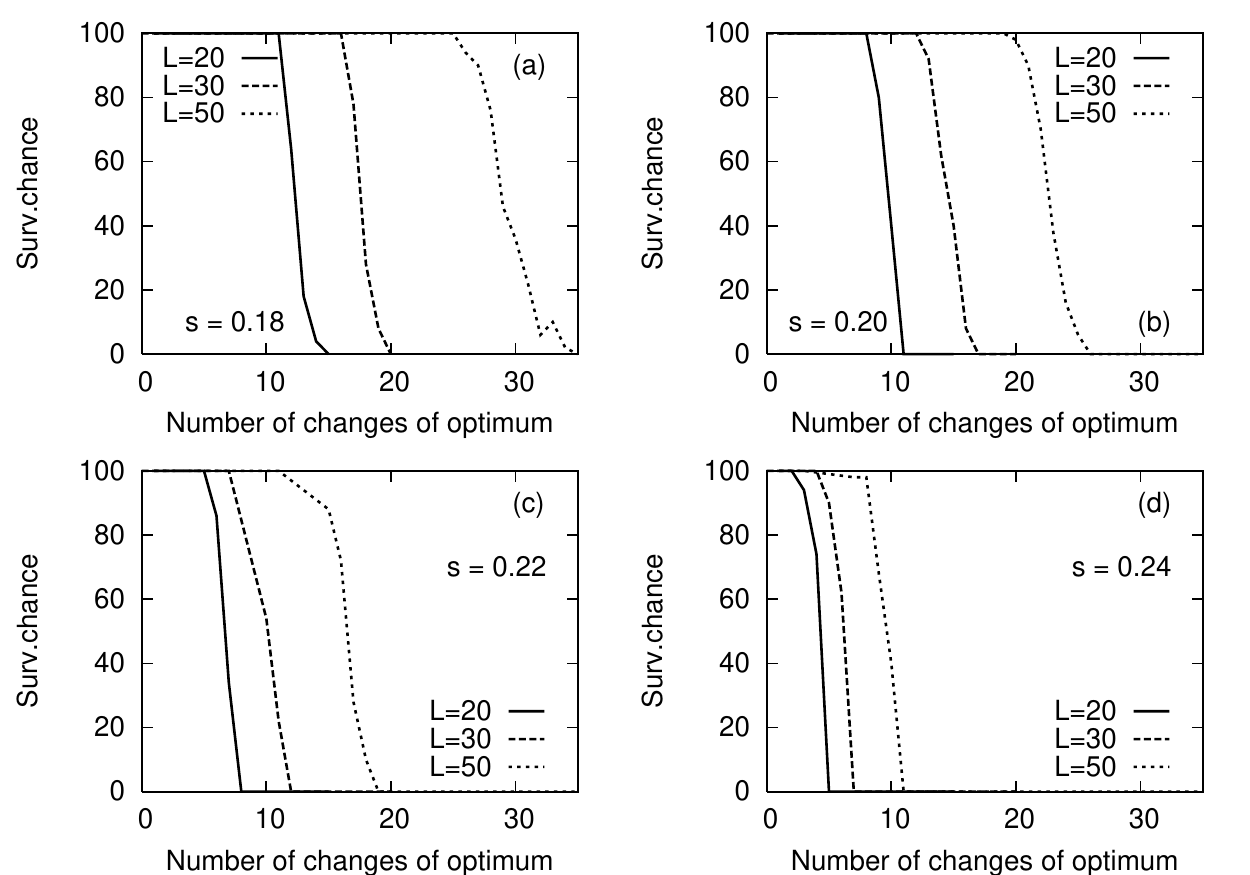}
\caption{Survival chance versus number of changed loci in the
optimum for short ($L$=20), medium ($L$=30) and long ($L$=50)
genotypes. Selection values are (a) $s$ =0.18, (b) $s$ = 0.20, (c)
$s$ = 0.22 and (d) $s$ = 0.24. $K$ = 2500.} \label{fig8}
\end{figure}

The survival chance of a population as a function of the number of
changes in the optimum (Figure \ref{fig8}) clearly depends on the
length of the genotype.

\begin{figure}
\centering
\includegraphics[scale=0.8]{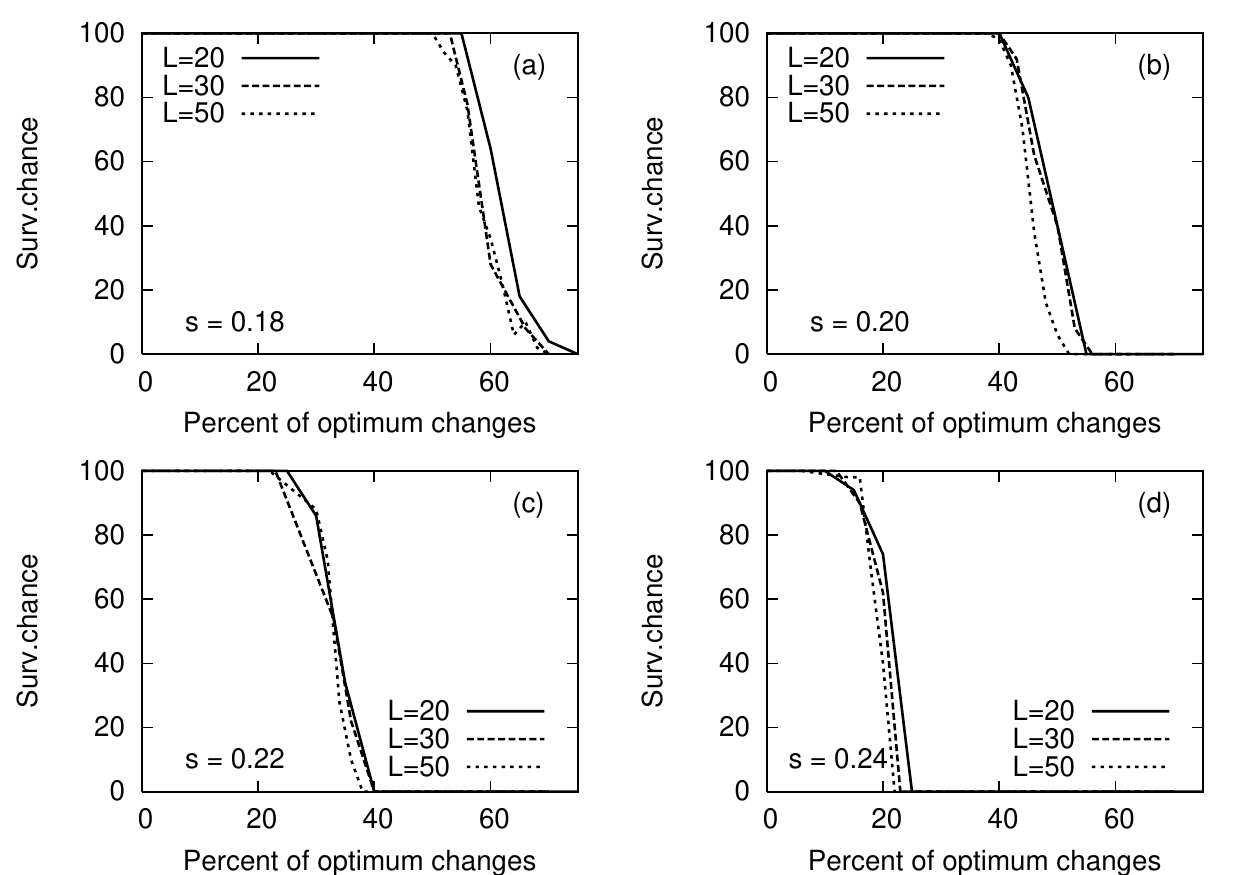}
\caption{The same data as in Figure \ref{fig8} except that now the
survival chance is plotted against the percentage of changed loci,
not their absolute number.} \label{fig9}
\end{figure}

If we however plot the survival chance against the relative change
in the optimum, i.e. the percentage of changes, then, as seen from
Figure \ref{fig9}, the differences between populations with
genotypes of different length disappear. Rather strong dependence on
the selection pressure has the same character as before.

\begin{figure}
\centering
\includegraphics[scale=0.5]{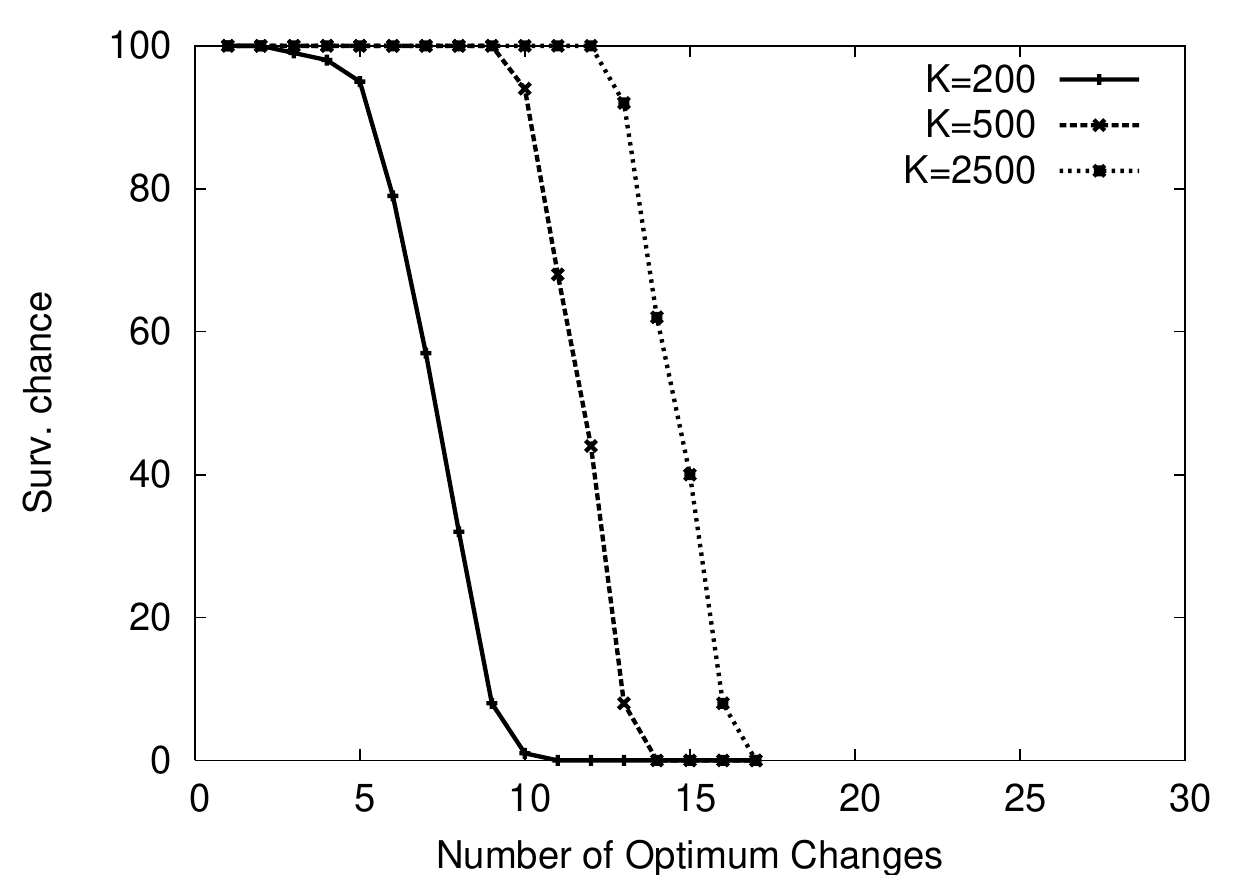}
\caption{Survival chance versus the number of changed loci for three
values of the carrying capacity -- $K$ = 200, $K$ = 500 and $K$ =
2500. $L$ = 30, $s$ = 0.20} \label{fig10}
\end{figure}

Figure \ref{fig10} shows that small populations ($K$ = 200) have a
lesser chance to survive medium or large scale catastrophes than
bigger populations. This is clearly different from what we have
found for periodic changes.

The threshold values of the selection, $s_c$, could be deduced from
Figures \ref{fig8} and \ref{fig10}, as the points where the curves
touch the horizontal axis. Hence, e.g. $s_c$ = 0.17 for $K$ = 2500
in Figure \ref{fig8}.
\section{Conclusions}
We have presented a model of population dynamics where two types of
the habitat changes are possible --  oscillatory ones (with
different periods of the oscillations) and abrupt ones, which may be
called catastrophes, when the scale of the catastrophe may vary.

As should be expected, selection pressure plays the crucial role in
each case, but when the changes are oscillatory, a small increase in
the value of the selection pressure shifts the population from the
''all survive`` into ''all extinct`` region. This effect is weaker
when the environment changes abruptly. Since the habitat after a
catastrophe remains unchanged, populations which survived it will
not decay, while in a periodically changing habitat surviving
initial oscillations is by no means a guarantee that a population
will also survive next changes, which, at the beginning, are from a
''better`` to a ''worse`` climate. Populations characterized by
longer genomes, presumably corresponding to more complex animals,
live longer in the case of periodic changes and could sustain bigger
catastrophes. Catastrophes are more dangerous for small than for
larger populations. Stochasticity plays a more important role in
small populations \cite{lande,shaffer,mdap}. As recently shown by
Shnerb e.a \cite{shnerb2}, life has a better chance on large
habitats. The situation is however different when the optimum is
oscillating. Here the size of the population, or more precisely, of
the carrying capacity, seems to have only small influence on the
fate of a population. If the selection pressure is strong enough in
the oscillating optimum, a small population will become extinct, but
this will happen most probably at the same time as for a large
population. In the case of a catastrophe, small populations face a
much bigger danger of being eliminated. If they however survive the
catastrophe they may live on, without a risk of elimination.

The results obtained by us, although using different simplifications
in construction of the models, agree with what has been found by
Shnerb e.a. \cite{shnerb} and biologists \cite{burger}  that a
population in conditions changing in time may either adapt  and live
well, or perish. Our finding that Hamming distance between the
optimum and the phenotype (bad gene) is not sufficient to predict
extinction, and some other, abiotic, factor influences the outcome,
corroborates the statement by Roberts and Newman \cite{roberts}. The
role of selection in the extinction probability has been, to the
best of our knowledge, not studied by physicists, although its
importance has been emphasized by biologists \cite{burger}.
Similarly, recombination, another very important factor in
diversification of the genetic pool  \cite{cebrat}, is often
neglected by physicists, apart from those dealing with the Penna
model (see e.g.\cite{stauffer2,cebrat2}).

There are several extensions of our model which could provide
interesting results  and determine the model robustness, like
changing the way  a genotype is transcribed to a phenotype.
Considering  the model on a lattice could tell what is the role
played by the topology of the system.  More realistic would be a
model with two sexes, where mating is possible only between
individuals of the opposite sex. This should be done on a lattice,
where the distance between the mates could play an important role.
Another question left open in this paper is how important is the
assumption that an individual could mate in each time step with a
different partner. How the results would change if the partners will
remain faithful to each other for all their lives?

\noindent
{\it Acknowledgment}\\
We are grateful to the anonymous referees for their valuable and helpful remarks.
MA admits some pertinent comments by P. Clippe.
AP acknowledges the support of the COST10 STSM  which permitted collaboration
with MA on this project. Preliminary stages were also supported via a Santander grant.
This work has been done within the framework of the UNESCO Chair of
Interdisciplinary Studies at the University of Wroc{\l}aw.

\end{document}